\documentstyle[12pt,twoside,fleqn,espcrc1,epsfig]{article}

\newcommand{\AmS}{{\protect\the\textfont2
  A\kern-.1667em\lower.5ex\hbox{M}\kern-.125emS}}
\newcommand{ \rts }{$\sqrt{s_{_{\rm NN}}}$ }
\newcommand{ \pt }{$p_t$ }
\newcommand{ \mpt }{$\langle p_t \rangle$ }

\newcommand{ \vv }{$\langle v_2 \rangle$ }

\title{Transverse Dynamics at RHIC} 
\author{Nu Xu
{\it Lawrence Berkeley National Laboratory, Berkeley, CA 94720, USA}
\\
 Zhangbu Xu
{\it Brookhaven National Laboratory, Upton, NY 11794 , USA}}

\begin{document}
\maketitle

\section{Initial Conditions}
Partonic collectivity refers to collective motion, i.e., hydrodynamic
type of flow, involving partonic degrees of freedom.  Large initial
energy deposition at central rapidity and subsequent strong
interaction at the partonic level is an important initial condition
for partonic collectivity. We first discuss what the measured average
transverse momentum distribution \mpt is.  The left panel of
Fig.~\ref{ptscaling} shows the \mpt of negatively charged particles
($h^{-}$) within $|\eta|<0.5$ or total charged particles ($N_{ch}$) as
a function of $\sqrt{s}$ for $pp,\bar{p}p$ (open circles),
$e^{+}e^{-}$(open triangles), and $AA$ (filled circles)
collisions~\cite{hminus,na49}.  For the $pp$ collisions the
dashed-line is the parameterization of the measured mean
$p_t$~\cite{ua1}. For the $e^{+}e^{-}$ data, the straight dashed-line
is an approximation of JETSET calculations~\cite{opal}.  The fact that
the \mpt from $AA$ collisions is distinctly different from both $pp$
and $e^{+}e^{-}$ indicates that the $AA$ collisions are not simple
superpositions of the elementary collisions. In heavy ion collisions,
the increase of \mpt as a function of beam energy or centrality is
much larger than expected from the soft-hard two-component model
\cite{tcm}. For example, the \mpt difference between SPS and RHIC is a
factor of 5 larger than the \mpt difference between these two energies
in $pp$ collisions.  In a simple two-component model, the maximum
difference is a factor of 3 if the increase in \mpt is solely due to
the increasing contribution from the hard scattering component.  The
right panel of Fig.~\ref{ptscaling} also shows the calculation from
RQMD with and without rescattering and HIJING which underpredicts the
\mpt.

\vspace{-0.75cm}

\hspace{-1.0cm}
\begin{minipage}{7.0cm}
  {\includegraphics*[width=4.5in]{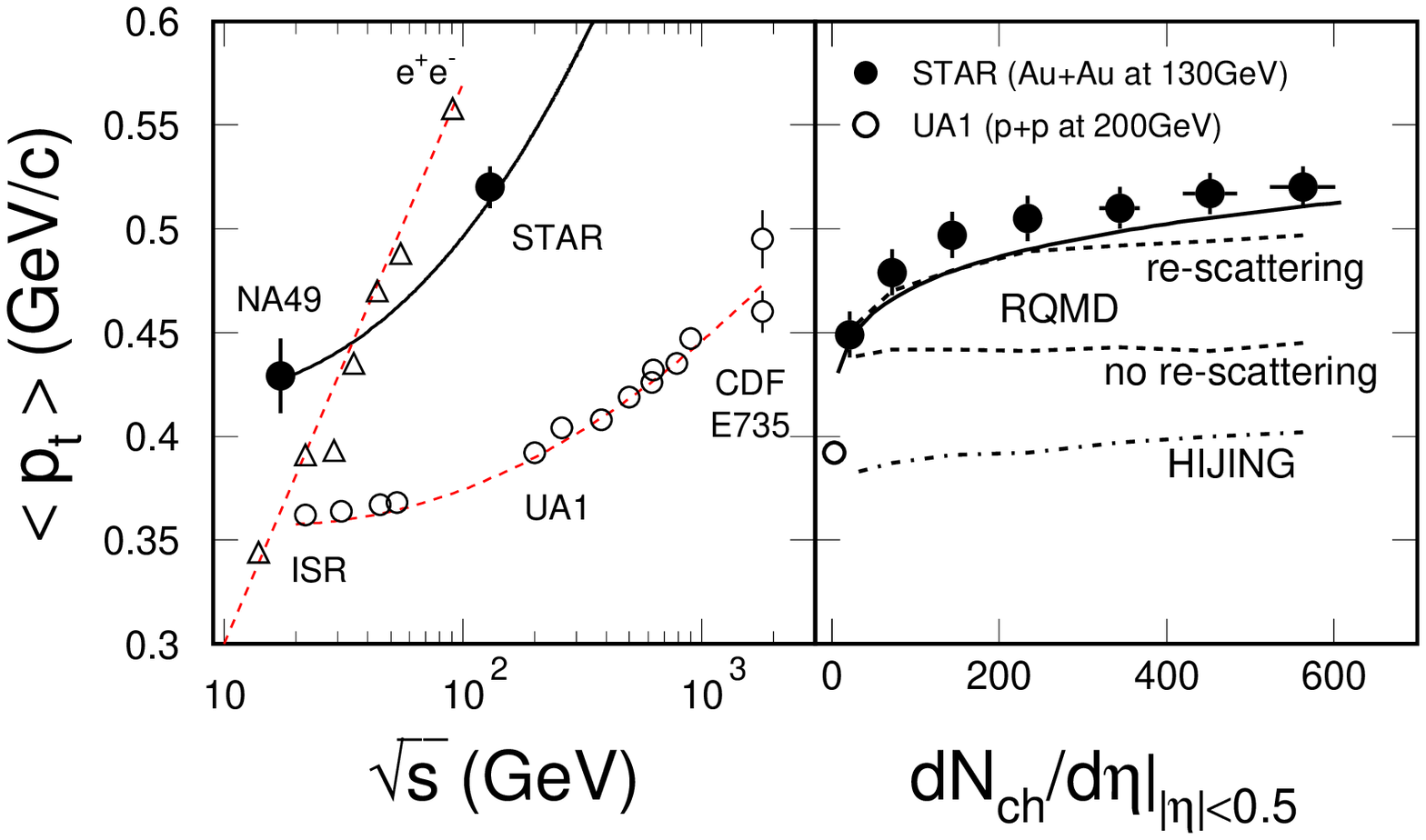}} 
\end{minipage}
\hspace{3.45cm}
\vspace{-5.75cm}
\begin{minipage}{5.45cm}
  \small{{\bf Figure 1 }({Left): Negatively charged hadron \mpt as a
      function of beam energy $\sqrt{s}$ for $e^+e^-, pp,\bar{p}p$
      (open symbols) and central $AA$ (filled symbols) collisions.
      (Right): \mpt as a function of collision centrality. The
      solid-curves represent \mpt scaling and the dashed ones
      represent RQMD with/without re-scatterings and the default
      HIJING(1.35) results.}}
  \label{ptscaling}
\end{minipage}

\vspace{5.5cm}

Both saturation~\cite{schaffner1} and hydrodynamic~\cite{hydroscaling}
models predict some scaling behavior of \mpt $\sim
\sqrt{({{dN}/{d\eta}})_{_{AA}}/{\pi{R^{2}}}}$ when the conditions are
satisfied.  The solid curves in Fig.~\ref{ptscaling} show the energy
dependence and centrality dependence of \mpt as discussed in
Ref.~\cite{xzb_moriond}.  However, the preliminary STAR
data~\cite{gene} show that the increase in \mpt from \rts = 130 to 200
GeV is not as strong as the thick-line indicates which may invalidate
the scaling law from the saturation model \cite{schaffner1}.  On the
other hand, the energy dependence might indicate the importance of
early thermalization in the partonic stage, as proposed by \mbox{A.\ 
  Mueller} {\it{et al.}}  \cite{mueller}. Such initial partonic
activity is consistent with the early development of flow as indicated
from \vv measurements at RHIC \cite{starflow}.  It will be interesting
to see if the description holds with an energy scan at lower beam
energies at RHIC.

\section{Partonic Collectivity at RHIC}

The transverse momentum distributions \cite{star,phelambda}, from
central Au + Au collisions at \rts = 130 GeV, are shown in Figure
\ref{kala2}. One can see that at \pt $\sim$ 2 GeV/c, all heavier mass
particle yields approach that of lighter ones. This behavior does not
depend on the particle type except their mass.  The mass dependence
indicates a strong transverse collective expansion in the collisions
at RHIC.  More importantly, from the azimuthal anisotropy \vv
\cite{starflow} combined with the spectra results, one might
conclude that the collective expansion has already developed at the
early partonic stage.

{\psfig{figure=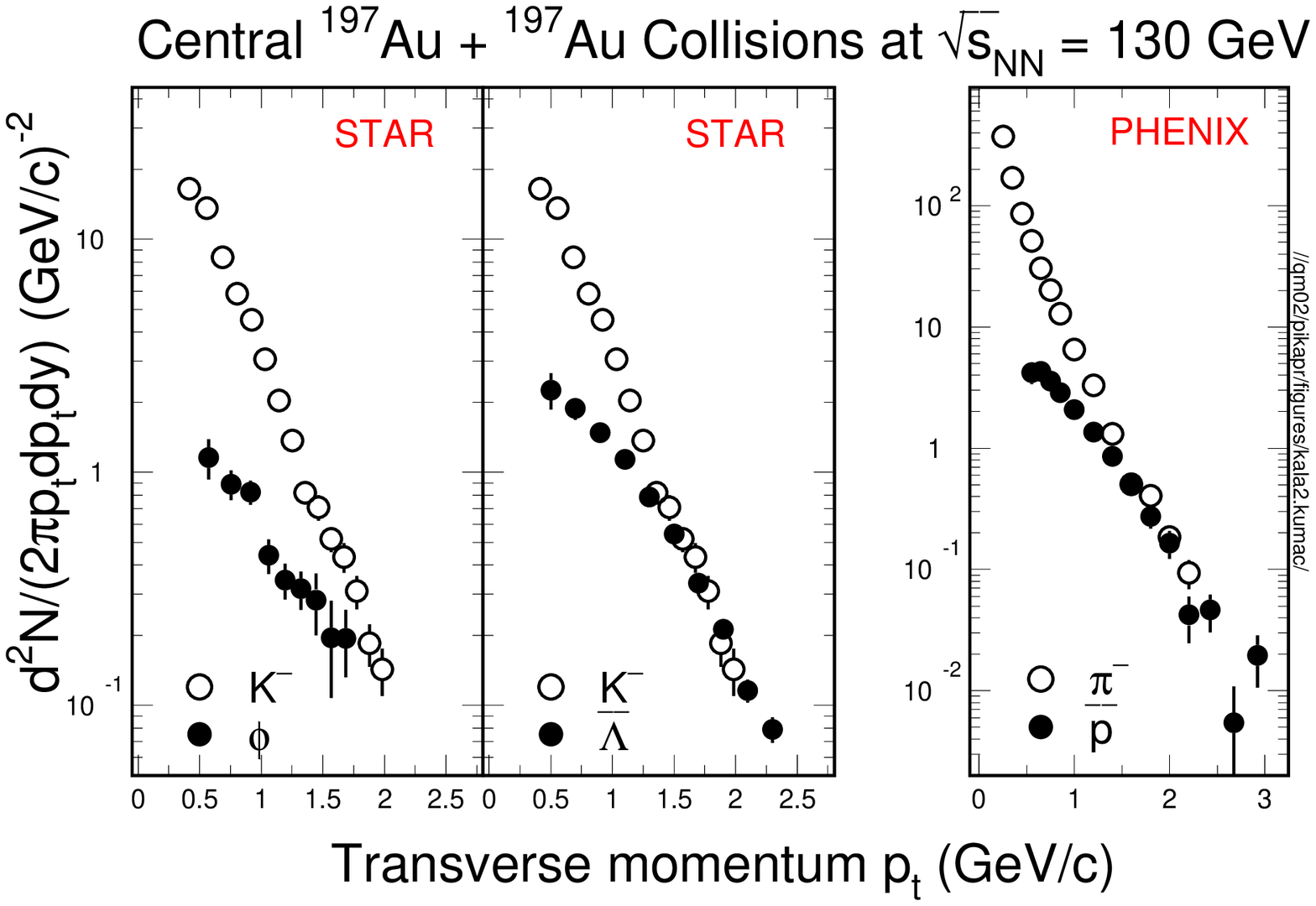,height=7.0cm}}

\vspace{-5.0cm}
\hspace{9.0cm}
\begin{minipage}{5.65cm}
  \small{{\bf Figure 2 }{Transverse momentum spectra from central
      Au+Au collisions at RHIC (\rts = 130 GeV). At $p_t \sim 2$
      GeV/c, heavier particle yields are approaching that of lighter
      ones - indicating a strong collective expansion developed in
      those collisions. }}
\label{kala2}
\end{minipage} \\

\vspace{1.0cm}

The measured transverse momentum distributions also have been fitted
by the function $f = A \cdot exp(-m_t/T),$ where $m_t =
\sqrt{p_t^2+m^2}$, $T$ is the inverse slope parameter, and $A$ is a
normalization constant.  The magnitude of the slope parameter provides
information about the temperature (random motion in local rest frame)
and collective transverse flow.  Fig.  3(left) shows the measured
particle slope parameters from Pb+Pb central collisions at \rts = 17.2
GeV ~\cite{na44_coll_exp,tsukuba,bordalo01,na50plb} (open symbols) and
from Au + Au central collisions at \rts = 130 GeV
\cite{star,phelambda,mauel01} (filled symbols).  As one can see in the
figure, at the SPS energy, the values of the slope parameters for the
$\Omega$ baryon, charm particles $J / \psi$, and $\psi'$
~\cite{bordalo01,na50plb} are all about 240 MeV independent of the
mass.  The straight hatched band indicates the possible variation of
the slope parameters due to initial scatterings \cite{hufner}. The
values of the slope parameters from other light hadrons like $\pi, K$,
and $p$, on the other hand, show a linear increase as the mass
increases.  Such a mass dependence of the slope parameter can be
explained by collective flow within the framework of hydrodynamics
\cite{na44_coll_exp,deark01,nxu01,heinz87}.
  
In the hadronic phase, the elastic scattering cross sections for
particles like $\phi, \Omega,$ and $J / \psi$ are smaller than that of
$\pi, K,$ and $p$~\cite{hsx98}.  Therefore, the interactions between
them and the rest of the system are weak, leading to the flat band
behavior in Fig.3(left).  On the other hand, the slope parameter of
these weakly interacting particles may reflect some characteristics of
the system at hadronization. Then it should be sensitive to the
strength of the color field~\cite{schwinger,soff,bleicher001}. Under
this assumption, the fact that the weak interacting particles show a
flat slope parameter as a function of particle mass would indicate
that the $flow$ develops at a later hadronic stage of the collision at
SPS.
  
In figure 3(right panel), the integrated values of $v_2$ demonstrate a
mass dependence as predicted by hydrodynamic calculations
\cite{starflow,pasi01}. In terms of hydrodynamics this mass dependence
is due to the anisotropic pressure gradient pushing massive particles
outward in velocity space. The width of the gray-band indicates the
uncertainties of the model calculations, mostly due to the choice of
the freeze-out conditions \cite{pasi01}. A key element in the
calculation is that the system develops the anisotropy at a relatively
early stage of the collision when partonic degrees of freedom dominate
the interactions. At higher collision energies, the increased
contribution from the partonic interactions lead to the observed
systematic increase of $v_2$ \cite{starflow}.
 
\vspace{-0.5cm}
{\psfig{figure=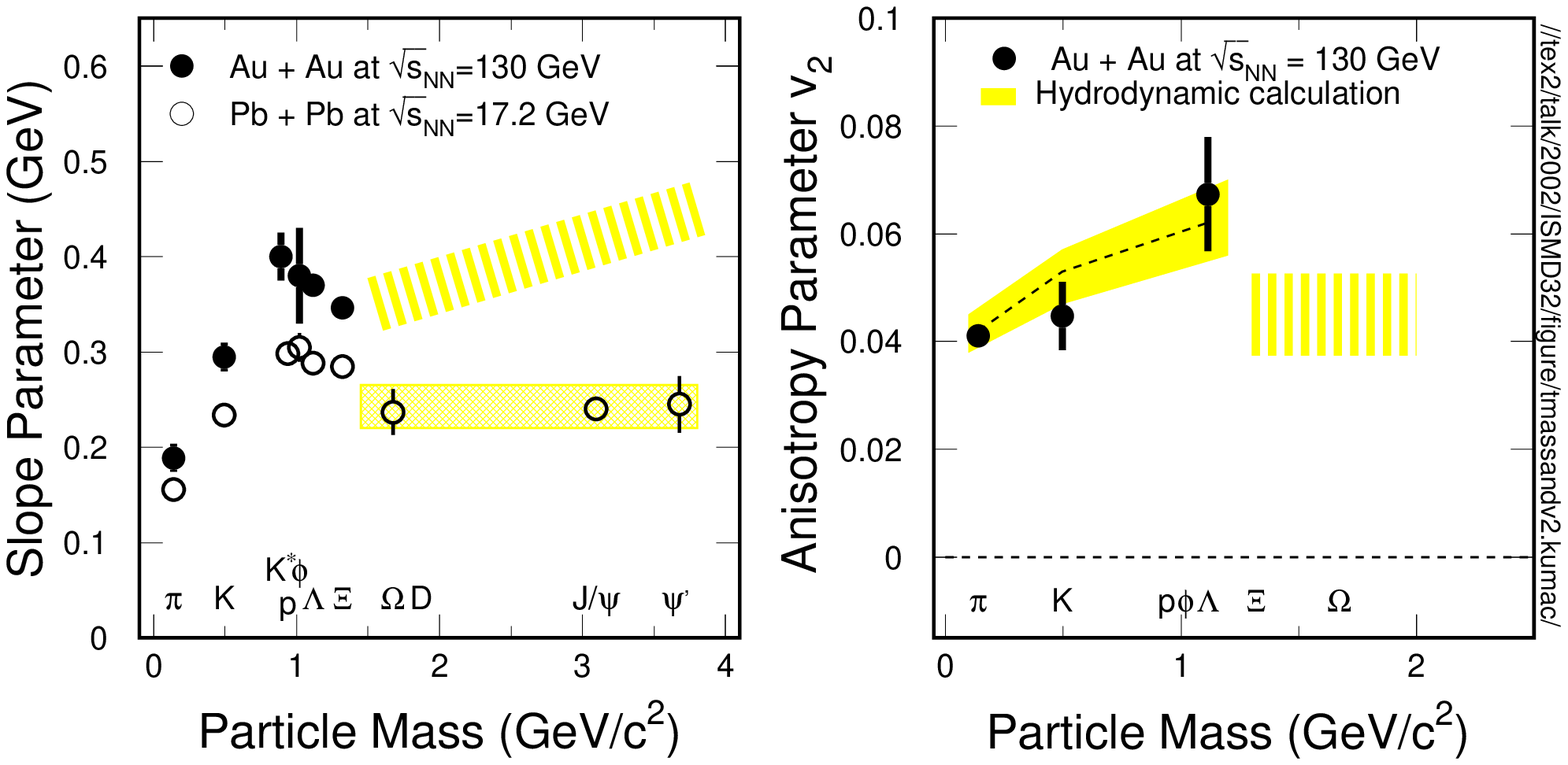,height=6.5cm}}

\vspace{-5.6cm}
\hspace{9.75cm}
\begin{minipage}{5.65cm}
  \small{{\bf Figure 3 }{ (Left) The slope parameter as a function of
      particle mass. Open and filled symbols represent the results
      from Pb+Pb and Au+Au central collisions at SPS and RHIC,
      respectively.  (Right) $v_2$ as a function of particle mass.
      The gray-band indicates the hydrodynamic model results
      \cite{pasi01}.}}
\label{tmassv2}
\end{minipage} 

\vspace{0.25cm}

At RHIC energies, see filled symbols in Fig.3(left), the mass
dependence of the slope parameter seems to be stronger than that
at SPS energies, indicating a larger collective
flow in higher energy nuclear collisions.  The strong energy
dependence of the slope parameters might be the consequence of the
larger pressure gradients at RHIC energies. With a set of reasonable
initial/freeze-out conditions and equation of state, the stronger
transverse expansion at RHIC energies was indeed predicted by
hydrodynamic calculations \cite{deark01,pasi01}.  Should the
collective flow develop at the partonic level, one would expect a mass
dependence of the slope parameters for particles like $\Omega$, $J /
\psi$, and $\psi'$ as indicated by the dashed band in Figure 3(left)
\cite{nxu01,pbmcharm,thews01}. For the same reason, one would
expect a non-zero value of $v_2$ for those particles as indicated by
the dashed band in the right panel of Fig.3.

\section{Summary}
Studies of $\langle p_t \rangle$, transverse momentum spectra and
anisotropy flow from nuclear collisions at RHIC indicate early
thermalization and strong collective expansion. We propose a
systematic study of the anisotropy parameter $v_2$ and the transverse
momentum distributions of $\phi, \Omega, D^0, \Lambda_C$ and $J /
\psi$ in order to extract information on partonic collective flow.

We are grateful for many enlightening discussions with Drs. P.
Braun-Munzinger, K. Haglin, P. Huovinen, D. Kharzeev, H.G.  Ritter,
J. Schaffner-Bielich, K.  Schweda, E.V.  Shuryak, R. Snellings, S.
Soff, D. Teaney, T. Ullrich, and E. Yamamoto.  This work has been
supported by the U.S.  Department of Energy under Contract No.
DE-AC03-76SF00098.



\end{document}